\documentstyle[emulateapj,psfig,apjfonts]{article}
\def\rxj{RX~J170849--400910}
\def\rxjsnr{G346.5--0.1}
\def\4u{4U~0142+61}
\def\1627{SGR~1627--41}
\def\etal{{\rm et~al.\ }}

\def\HII{H\,{\sc ii}}
\def\kms{km~s$^{-1}$}

\lefthead{GAENSLER ET AL}
\righthead{AXPS AND SGRS IN SUPERNOVA REMNANTS}

\begin{document}
\title{Anomalous X-ray Pulsars and Soft Gamma-Ray Repeaters
in Supernova Remnants}
\submitted{Accepted to {\em The Astrophysical Journal}\ on 2001 June 4}
\author{B. M. Gaensler\altaffilmark{1,5}, P. O. Slane\altaffilmark{2},
E. V. Gotthelf\altaffilmark{3} and G. Vasisht\altaffilmark{4}}
\altaffiltext{1}{Center for Space Research, Massachusetts Institute of
Technology, 70 Vassar Street, Cambridge, MA 02139; bmg@space.mit.edu}
\altaffiltext{2}{Harvard-Smithsonian Center for Astrophysics, 
60 Garden Street, Cambridge, MA 02138}
\altaffiltext{3}{Columbia Astrophysics Laboratory, Columbia University, 550
West 120th Street, New York, NY 10027}
\altaffiltext{4}{Jet Propulsion Laboratory, California Institute of
Technology, 4800 Oak Grove Drive, Pasadena, CA 91109}
\altaffiltext{5}{Hubble Fellow}

\begin{abstract}

Important constraints on the properties of the Anomalous X-ray Pulsars 
(AXPs) and Soft Gamma-Ray Repeaters (SGRs) can be provided
by their associations with supernova remnants (SNRs). We have made
a radio search for SNRs towards
the AXPs~\rxj\ and \4u\ --- we find that the former lies near a possible new 
SNR with which it is unlikely to be physically associated, but see
no SNR in the vicinity of the latter. We review all claimed pairings
between AXPs and SNRs; the three convincing associations imply that
AXPs are young ($<$10\,000~yr) neutron stars with low projected
space velocities ($<$500~\kms). Contrary to previous claims,
we find no evidence that the density of the ambient medium
around AXPs is higher than that in the vicinity of radio pulsars.
Furthermore, the non-detection of radio emission from AXPs
does not imply that these sources are radio-silent.

We also review claimed associations between SGRs and SNRs. We find none
of these associations to be convincing, consistent with a scenario in
which SGRs and AXPs are both populations of high-field neutron stars
(``magnetars'') but in which the SGRs are an older or longer-lived
group of objects than are the AXPs.  If the SGR/SNR associations are
shown to be valid, then SGRs must be high-velocity objects and most
likely represent a different class of source to the AXPs.

\end{abstract}


\section{Introduction}
\label{sec_intro}

It has long been believed that radio pulsars are the most common
form of young neutron star. However, a variety of new discoveries
have demonstrated that the population of such objects is
considerably more diverse than originally thought.
Most prominent amongst these recent results
has been the emergence of two possibly related classes of source,
the Anomalous X-ray Pulsars (AXPs) and the Soft Gamma-Ray Repeaters
(SGRs). 

The AXPs are a group of 5--6 unresolved X-ray sources, showing
pulsations at periods of 6--12~sec and generally exhibiting steady
spin-down (see \cite{mer00} for a comprehensive review).
None of the AXPs show any evidence for a binary companion,
and none have been detected at radio wavelengths.
The AXPs all lie in the Galactic Plane, and several are located near
the centers of supernova remnants (SNRs).

The four (possibly five) SGRs are characterized by their
occasional intense $\gamma$-ray activity (see \cite{hur00b} for a
review).  All four confirmed SGRs have been identified with persistent
X-ray sources. In three cases, pulsations have been detected from
these X-ray sources, with periods in the range 5--8~sec and positive
period derivatives.
Radio pulsations have been claimed from one SGR 
but have not been confirmed (\cite{spk00}; \cite{lx00}).  
Three of the SGRs are in
the Galactic Plane while a fourth is located in the Large Magellanic
Cloud (LMC); in all four cases, the SGR lies near a cataloged SNR.

The generally accepted interpretation for the SGRs is that they
are ``magnetars'' --- young isolated neutron stars with extremely high
($\ga10^{15}$~G) magnetic fields (\cite{pac92};
\cite{dt92a}; \cite{kds+98}). Such field strengths are
well above the quantum critical field limit of $4\times10^{13}$~G;
the physics and emission processes of the SGRs are therefore expected
to differ considerably from those of lower-field radio pulsars
(\cite{dun00}).  These high magnetic fields can account
for the $\gamma$-ray activity seen from the SGRs,
and also for their observed periods and period-derivatives.

An interpretation for the AXPs is less clear. Given the similarities
of their timing properties to those
of the SGRs, several authors have
proposed that the AXPs are also magnetars (\cite{td96b};
\cite{vg97}) --- the dipole magnetic
fields inferred from their periods and period-derivatives are
indeed in the range $10^{14}-10^{15}$~G. Because $\gamma$-ray burst
activity has not been observed in any of the AXPs, it has been suggested
that they represent a different evolutionary stage than do SGRs in the life 
of a magnetar (\cite{kds+98}; \cite{gvd99}; \cite{ggv99}).

Alternatively,
a variety of authors have argued that the AXPs are neutron
stars with typical magnetic fields $\sim10^{12}-10^{13}$~G,
and whose X-ray flux originates from accretion from
a low mass
companion (\cite{ms95}), a supernova fall-back disk (\cite{chn00}), or
a common-envelope phase of evolution (\cite{vtv95}).  
These interpretations can better explain the small range of periods
observed for the AXPs (\cite{ch00}), but have trouble
accounting for these sources' faint optical counterparts (\cite{hvk00}).

An important constraint in interpreting both AXPs and SGRs
has been the associations of these sources with SNRs. Such an
association confirms that the AXP or SGR was formed in a supernova
explosion and is thus likely to be a neutron star. Furthermore,
since SNRs are reasonably short-lived (20--100~kyr; \cite{bgl89};
\cite{sfs89}), the
association argues that the AXP/SGR is comparatively young. 
Beyond these basic conclusions, an association with a SNR also provides
an estimate of a neutron star's age, distance, and birth-site,
from the latter of which its projected space velocity can be estimated.
Associations with SNRs have correspondingly been used to infer
a variety of properties for the AXPs and SGRs: it has
been concluded from SNR associations that SGRs have
extremely high space velocities (\cite{td95}), that AXPs are
very young objects (\cite{vg97}), that
AXPs and SGRs are born in higher density regions
than are radio pulsars (\cite{mlrh01}),
that the supernova explosions which produce SGRs
occur on the edge of their progenitor's wind-bubbles
(\cite{gav01}),
and even that AXPs and SGRs are strange stars (\cite{dd00};
\cite{zxq00}).

It is clearly of considerable interest to identify new instances of
such associations, but at the same time it is crucial to determine
whether previously-claimed associations are genuine. There are many
regions of the Galactic Plane where the spatial density of SNRs is high. The
possibility that an AXP/SGR and an adjacent SNR are physically
unrelated and merely lie along similar lines-of-sight must therefore be
carefully considered.  Indeed, many associations claimed between SNRs
and radio pulsars (\cite{kw90}; \cite{car93}; \cite{mop93}) have
subsequently been shown to be spurious (\cite{fkw94}; \cite{njk96};
\cite{sgj99}), and similar caution must be applied to associations
between SNRs and AXPs/SGRs. 

Associations between SNRs and radio pulsars are usually judged on
criteria such as agreement in ages and distances and whether the
transverse velocity implied for the pulsar is reasonable.  However,
these criteria are problematic when applied to cases involving AXPs and
SGRs.

The characteristic age parameter
used to estimate the ages of radio pulsars ($\tau \equiv P/2\dot{P}$, where
$P$ is the spin-period and $\dot{P}$ is the period-derivative)
is applicable only if a neutron star's spin-down is entirely due
to magnetic dipole radiation. However, in the case of a magnetar
there is expected to be a significant additional torque
due to a relativistic particle wind (\cite{tb98}; \cite{hck99}).
Furthermore, some AXPs and SGRs show complicated timing
behavior not consistent with any kind of steady spin-down
(\cite{wkv+99b}; \cite{kgc+01}).
Finally, in the specific case of the otherwise
convincing association  of
the AXP~1E~2259+586 with the SNR~CTB~109 (discussed
in Section~\ref{sec_axp_other}), the age of the SNR
is a factor $\sim$20 times smaller than the value
of $\tau=P/2\dot{P}$ inferred for the AXP.
There thus seems to be neither any evidence nor expectation
that the ages of AXPs and SGRs can be usefully estimated from their
observed spin parameters. 

Distance estimates for most of the AXPs/SGRs come only
from a measurement of the column density of absorbing material
along the line-of-sight, and so have large uncertainties $>$50\%.
Finally, while transverse velocities inferred for radio pulsars
can be compared to the pulsar velocity distribution  as
determined from proper motion measurements
(e.g.\ \cite{fgw94}), nothing is known
about the range of velocities expected for AXPs and SGRs.
Therefore the only criterion we can reliably apply in the case of AXPs
and SGRs is positional coincidence --- namely,
whether the neutron star and SNR are sufficiently close on the sky
that it is unlikely that they align by chance (e.g.\ \cite{kf93};
\cite{sbl99}). 

Motivated by these considerations, we here present a study of associations
of AXPs and SGRs with SNRs, a brief discussion of which
was outlined by Gaensler (2000\nocite{gae00}).
In Sections~\ref{sec_obs} and \ref{sec_results} we describe a search for 
radio SNRs towards two AXPs, \rxj\ and \4u.
In Section~\ref{sec_axps} we discuss
the implications of these observations, review possible SNR associations
for all other known AXPs, and use the results to infer some general
properties about the AXP population. In Section~\ref{sec_sgrs} we consider the
case of SGRs in SNRs, similarly derive some overall properties of the
SGR population, and discuss the possible relationship between AXPs and SGRs.

\section{Observations and Data Reduction}
\label{sec_obs}

The regions surrounding the AXPs \rxj\ and \4u\ were observed
with the Very Large Array (VLA), and are summarized in Table~\ref{tab_obs}.
All observations were centered at 1435~MHz, using
three adjacent frequency channels each of width 12.5~MHz.
Each field was observed in a 3-point mosaic pattern, both to increase
the field-of-view and to maximize sensitivity to extended structure.
Antenna gains and polarization were calibrated using regular
observations of PKS~B1827--360 (for \rxj) and
3C~468.1 (for \4u).
Amplitudes were calibrated using observations of 3C~286 (for \rxj)
and 3C~48 (for \4u), assuming flux densities at 1.4~GHz
of 15.0~Jy and 16.1~Jy respectively. 

Data were analyzed using the {\tt MIRIAD}\ package. Once visibilities
were edited and calibrated, an image corresponding to the 
combination of all three frequency
channels was made using multi-frequency synthesis (\cite{sw94}).
All three pointings towards each source were then deconvolved
simultaneously
using the {\tt MOSMEM}\ algorithm (\cite{ssb96}). The resulting
image was then smoothed with a gaussian restoring beam of
dimensions corresponding to the diffraction-limited resolution
of the observations.

Our VLA data on \4u\ lacks sensitivity to the largest spatial
scales. To search for such emission,
we have also obtained a 1.4-GHz image of the region
surrounding \4u\ made 
as part of the Canadian Galactic Plane Survey (CGPS; \cite{eti+98})
and archived by the Canadian Astronomy Data Center. This image
includes both interferometric and single-dish data, and so has
sensitivity to all scales down to the resolution limit
of 1~arcmin.

\section{Results}
\label{sec_results}

\subsection{\rxj}

\subsubsection{Imaging}
\label{sec_results_rxj_image}

A 1.4~GHz image of the field surrounding \rxj\ is shown
in Fig~\ref{fig_rxj}. Only the lower-resolution DnC-array data
have been used to make this image, so as to give maximal
surface-brightness sensitivity.

It is clear that \rxj\ is in a complex region. Previously
identified sources which can be seen in the image include
the SNR~G346.6--0.2 ($\sim15'$ to the east of the AXP; \cite{dmgw93}), 
the ultra-compact \HII\ region G346.52+0.08 $=$~IRAS~17052--4001
($4'$ to the north of the AXP; \cite{wbhr98}),
the unresolved radio source G346.472+0.053 
($1'$ to the west; \cite{zhb+90}), and
the thermal source IRAS~17056--4004 ($4'$ to the east).

Two~arcmin to the east of the AXP can be seen an arc of diffuse
emission, running from north to south for $\sim8'$ before curving
around to the east, eventually fading and merging with
SNR~G346.6--0.2.  The overall morphology is that of a faint
partial shell, with a radius of $\sim6'$.
We have looked for this source in various archival
data-sets:  1.4~GHz VLA observations of SNR~G346.6--0.2 (\cite{dmgw93}),
1.4~GHz observations of SNR~G347.3--0.5 using the
Australia Telescope Compact Array (\cite{esg01})
and in the 843~MHz Molonglo Galactic Plane Survey (MGPS; \cite{gcl98}).
Although these observations are all of poorer sensitivity
than the data presented here, the presence and morphology of
this source is clearly apparent in all three data-sets.  In future
discussion, we refer to this source as \rxjsnr, corresponding to
the position of its approximate center of curvature.

An examination of the {\em IRAS}\ Galaxy Atlas (\cite{ctpb97}) shows
no counterpart to \rxjsnr\ at wavelengths of 60~$\mu$m or 100~$\mu$m,
nor in a map of the ratio 60~$\mu$m/100~$\mu$m. We also have
found no X-ray emission from this source in archival {\em ROSAT}\
or {\em ASCA}\ data.

We have searched for radio emission from \rxj\ itself by
making a 1.4~GHz image using only data from the higher-resolution
CnB array. No emission is seen at the position of \rxj\
in this image down to a $5\sigma$ upper limit of 3~mJy.

\subsubsection{Spectral Index Determination}

\begin{figure}[htb]
\vspace{3cm}
\centerline{\psfig{file=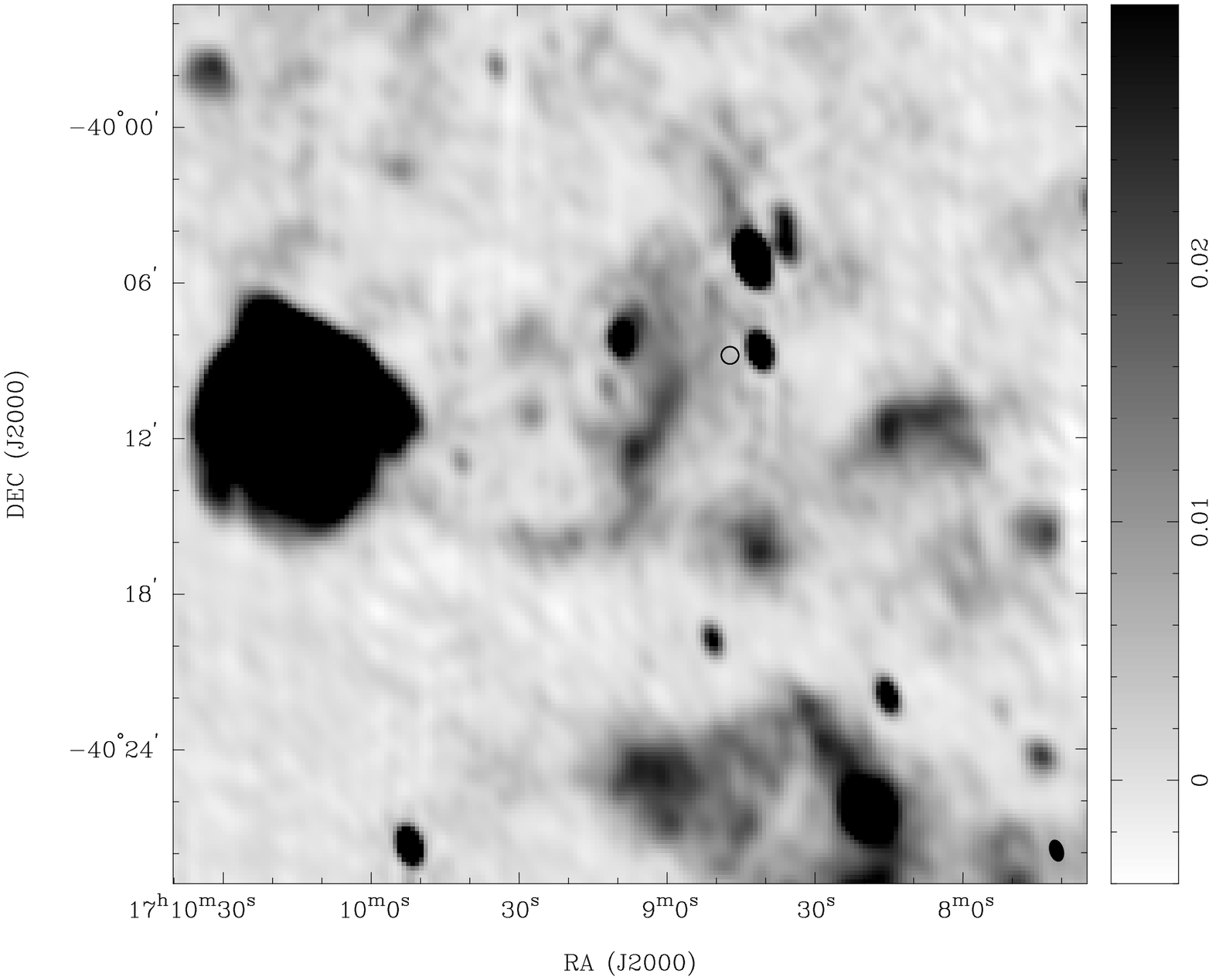,width=7in,angle=270}}
\caption{1.4~GHz VLA image of the field surrounding \rxj.  The
synthesized beam (shown at the lower right) is $52''\times33''$
and the RMS  noise is 0.7~mJy~beam$^{-1}$.  The greyscale ranges
between --4 and +30~mJy~beam$^{-1}$, as shown by the wedge to the
right of the image.  The position of \rxj, as given by Israel \etal\
(1999\protect\nocite{ics+99}), is marked by a circle (the positional
uncertainty is half the size of the circle). The bright extended source
at the eastern edge of the figure is the SNR~G346.6--0.2.}
\label{fig_rxj}
\end{figure}

We have determined approximate spectral indices for the
emission seen in Fig~\ref{fig_rxj} by comparing our 1.4~GHz
image to a 843~MHz MGPS image of the same region (\cite{gcl98}). 
In order to make a proper comparison of the images, 
we first applied to the 843~MHz data 
the mosaicing pattern, primary beam attenuation and $u-v$ coverage of the
1.4~GHz VLA observations so as to produce two data-sets which were
identical except in their brightness distributions. We 
then applied to this spatially-filtered 843~MHz data the same
deconvolution process as was required to produce the
1.4~GHz image shown in Fig~\ref{fig_rxj}.
Images at both frequencies were then smoothed to a resolution
of $60'' \times 60''$, and compared using the technique of spectral tomography,
whereby scaled versions of the 843-MHz
image are subtracted from the 1.4-GHz image, and the scaling factor
for which a given feature merges into the background is used to compute
that feature's spectral index (\cite{kr97}).
An accurate spectral index for \rxjsnr\ is
difficult to determine because it is so faint and diffuse. However, for
the brightest region along its extent, we estimate a spectral index
$\alpha = -0.4\pm0.2$ (where $S_\nu \propto \nu^\alpha$).

Of the other main sources seen in Fig~\ref{fig_rxj},
we find that SNR~G346.6--0.2 has a spectral index $\alpha = -0.6\pm0.1$,
in good agreement with the value $\alpha = -0.5$
tabulated by Green (2000\nocite{gre00}) in his SNR catalog.
G346.52+0.08 and IRAS~17056--4004 both have relative flat spectra
($-0.15\pm0.1$ and $-0.1\pm0.1$ respectively), consistent with their
interpretations as thermal sources. G346.472+0.053 has a steeper
spectrum, $\alpha = -0.9\pm0.1$, indicating that it is probably a
background radio galaxy.  

\subsection{\4u}

\begin{figure}[htb]
\vspace{3cm}
\centerline{\psfig{file=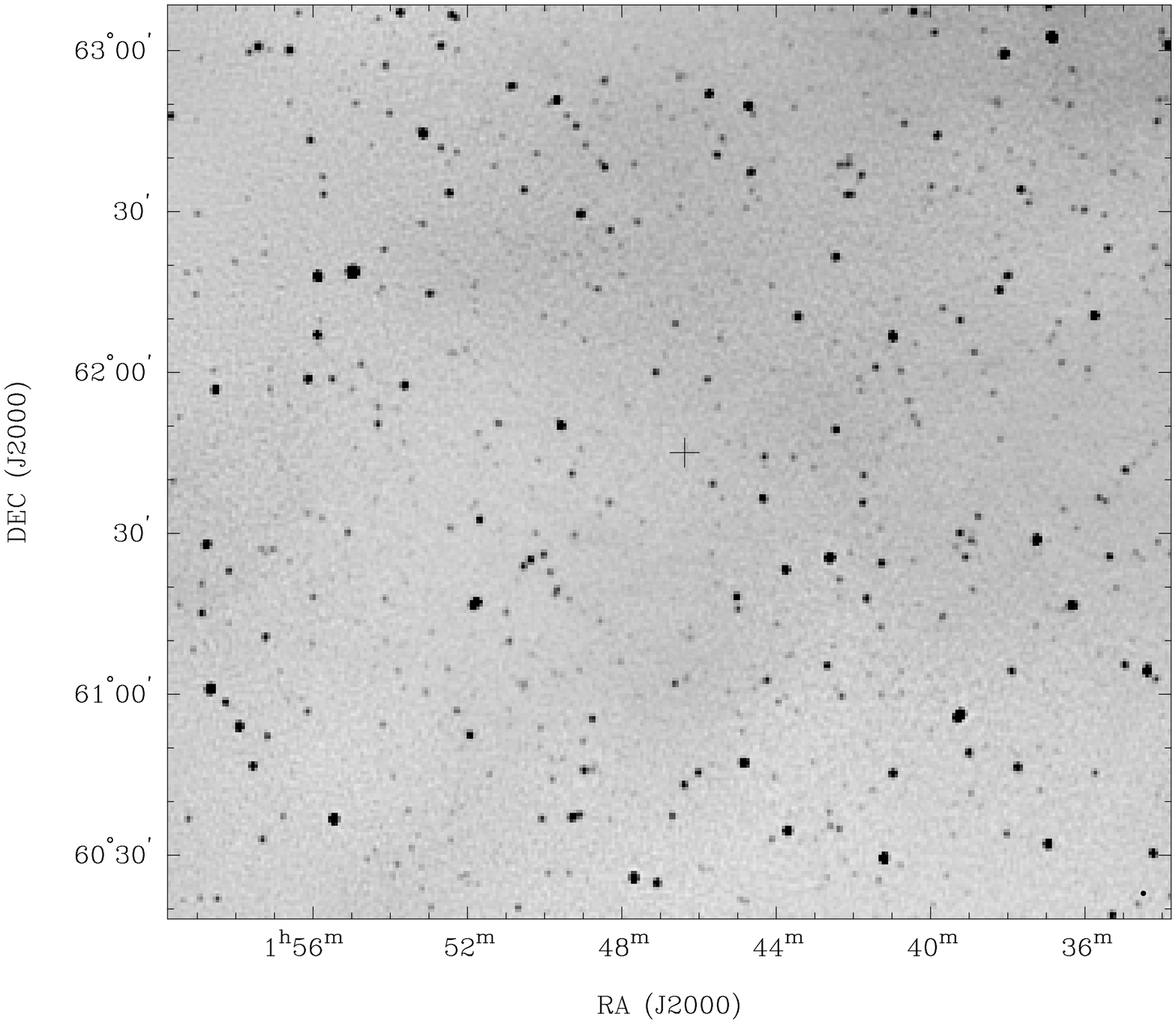,width=7in,angle=270}}
\caption{1.4~GHz CGPS image of the field surrounding \4u.  
The synthesized beam (shown at the lower right) is $60''\times60''$
and the RMS noise is 0.15~mJy~beam$^{-1}$. The greyscale ranges from
+9 to +15~mJy~beam$^{-1}$. The position of the optical
counterpart of \4u, as given by Hulleman, van Kerwijk \& Kulkarni
(2000\protect\nocite{hvk00}), is marked by a cross, and has
negligible positional uncertainty on this scale.}
\label{fig_4u}
\end{figure}

The CGPS image of the region surrounding \4u\ is shown in
Fig~\ref{fig_4u};
no extended structure can be seen anywhere near the position
of the AXP. The RMS sensitivity of the image is 0.2~mJy~beam$^{-1}$.
Assuming a typical SNR spectral index $\alpha = -0.5$, this
corresponds to a $1\sigma$ surface brightness limit on any SNR
of $\Sigma_{\rm 1\, GHz} = 3.5 \times 10^{-23}$~W~m$^{-2}$~Hz$^{-1}$~sr$^{-1}$.

We can use our VLA observations to search for emission from \4u
itself. Using these data, we find no point source at the position of \4u\
down to a $5\sigma$ limit of 0.27~mJy.

\section{Associations of AXPs with SNRs}
\label{sec_axps}

\subsection{\rxj}
\label{sec_axps_rxj}

The arc of emission \rxjsnr\ which we have identified in
Fig~\ref{fig_rxj} has a partial-shell morphology, a high radio/IR
flux ratio, and a spectral index which suggests non-thermal
emission. We thus suggest that \rxjsnr\ is potentially a
previously-unidentified SNR, although confirmation of this possibility
will require a detection of linear polarization and/or a more accurate
determination of its spectral index. It is certainly not unusual
for deep imaging to reveal previously-unidentified
faint SNRs in complex regions of the
Galactic Plane (\cite{fgw94}; \cite{ggv99}; \cite{cgk+00}) --- these
results underscore the incompleteness of current SNR samples.

Of the nearby sources in the field,
SNR~G346.6--0.2 is at a distance of either 5.5 or 11~kpc
(\cite{kfg+98}), while the ultra-compact \HII\ region G346.52+0.08
is at a distance of $\sim$17~kpc (\cite{ch87b}; \cite{cvew+95}).
Thus even if the case can be made that \rxjsnr\ is associated
with one of these nearby sources, its distance is still highly uncertain.
We assume in future
discussion that the distance to \rxjsnr\ is $10d_{10}$~kpc
with $0.5 \la d_{10} \la 2$.

If \rxjsnr\ is indeed a SNR, could it be associated with \rxj?
The hydrogen column density towards the latter as inferred from its
X-ray spectrum 
suggests that $d_{10} \ga 1$ for the AXP (\cite{snt+97}). So while
the distance to neither object is known, there is nothing to
suggest they are inconsistent. 

If we assume that the AXP was born at the center of \rxjsnr\ $10^4t_4$
years ago, we can infer a projected velocity for the AXP of
$2300d_{10}/t_4$~\kms.  If we assume
$t_4 \ga 5$ (as suggested by the faint 
and ragged appearance of \rxjsnr\ and the lack
of any X-ray emission from it)
and $d_{10} = 1$, we obtain a projected velocity of $\la460$~\kms,
which is comparable to that inferred for other AXPs
(see discussion in Section~\ref{sec_axp_overall} below).

However, we also need to consider the possibility that the AXP and
candidate SNR lie near each other only through random alignment along
the line-of-sight.  We can estimate such a probability by noting that
there are 12 SNRs in the catalog of Whiteoak \& Green (1996\nocite{wg96})
within a representative area bounded by $340^\circ \le l \le 350^\circ$,
$|b| \le 0\fdg5$.  To calculate the probability of an AXP
lying within $2'$ of the rim of an unrelated SNR, we inflate
the radius of each of these 12 SNRs by $2'$.
The chance of an alignment is then just the ratio of
the area of these 12 sub-regions to the total 10 square degrees
under consideration (\cite{sbl99}), corresponding to a probability
of $\sim$5\%. However, this
value is certainly an underestimate, as  we have carried out a targeted
observation of greater sensitivity than the survey of Whiteoak \& Green
(1996\nocite{wg96}), and the probability of finding a nearby SNR will
thus be higher than inferred from this catalog. The catalog of Whiteoak
\& Green (1996\nocite{wg96}) is complete to a 1-GHz surface brightness
of $\Sigma \approx 8\times 10^{-21}$~W~m$^{-2}$~Hz$^{-1}$, about 10
times poorer than the 3$\sigma$ sensitivity of the VLA observations
presented here. Gaensler \& Johnston (1995\nocite{gj95c}) simulate the
Galactic SNR population, and find that for a search 10 times deeper than
that of Whiteoak \& Green (1996), one would find approximately twice as
many SNRs.  
Similarly, Helfand \etal\
(1989\nocite{hvbl89}) argue that about 50\% of SNRs in this direction
are still to be discovered.

We thus conclude that for our targeted search, the probability of
finding a random alignment of the AXP with a nearby SNR is $\sim$10\%.
This does not imply that there is a $\sim$90\%
probability that \rxjsnr\ and AXP~\rxj\ are physically associated, but
simply that there is a 90\% probability that the AXP is not randomly
located with respect to the SNR distribution. If AXPs are young neutron
stars, they will be naturally clustered into the same small regions of
the sky as SNRs. A low probability of random alignment may simply
indicate
that AXPs and SNRs are both associated with massive star formation.  We
therefore conclude that there is no compelling argument for a physical
association between \rxjsnr\ and \rxj.

As an aside, we note that Marsden \etal\ (2001\nocite{mlrh01}) have
proposed an association between \rxj\ and SNR~G346.6--0.2. The AXP
is $12'$ beyond the perimeter of this SNR, and 
the probability of random alignment (estimated through the same
arguments as above) is $\sim$30\%. We therefore
consider it highly unlikely that there is any association
between \rxj\ and G346.6--0.2.

\subsection{\4u}

There is no evidence for any SNR towards \4u\ down to a 
1-GHz brightness limit 
$\Sigma = 3.5 \times 10^{-23}$~W~m$^{-2}$~Hz$^{-1}$~sr$^{-1}$.

The faintest SNRs in the Galactic SNR catalog are
G65.2+5.7 and G156.2+5.7 (\cite{gre00}). In both cases,
most of the extended emission from the SNR shell
has a brightness of at least $\Sigma_{\rm 1\, GHz} \ga
1.5\times10^{-22}$~W~m$^{-2}$~Hz$^{-1}$~sr$^{-1}$, which
would have been easily detected in the observations presented here.
Thus if there is a SNR associated with \4u, it is fainter
than any other radio SNR known. 

Recent X-ray results
suggest that some SNRs are detectable at
high energies but have no corresponding radio emission
(e.g.\ \cite{scs+00}); however no spatial or spectral
evidence of a SNR has been reported in X-ray observations
of this source (e.g.\ \cite{wae+96}).

\subsection{Other AXPs}
\label{sec_axp_other}

There are three (possibly four) other AXPs, several of
which lie within SNRs. We now briefly describe each case.

The AXP 1E~2259+586 lies within $3'$
of the geometric center of the young SNR~CTB~109 (G109.1--1.0;
\cite{fg81}).
In a representative area $100^\circ < l < 120^\circ$, $|b| < 2^\circ$,
there are just five SNRs listed
in Green's (2000\nocite{gre00}) catalog. 
The probability of the AXP lying this close
to the center of one of these SNRs by chance is $5\times10^{-4}$. 
This probability is sufficiently low that we think
it highly probable that 1E~2259+586
and CTB~109 form a genuine association.

Similarly, AXP~1E~1841--045 lies $\la30''$ from the center of the
very young SNR~G27.4+0.0 (Kes 73; \cite{vg97}). Considering
the 11 SNRs in the region $20^\circ < l < 30^\circ$, $|b| < 1^\circ$, 
we calculate the probability of chance alignment
to be $1\times10^{-4}$. As for 1E~2259+586 / CTB~109,
it seems highly likely that 1E~1841--045 and SNR~G27.4+0.0 form a
physical association.

AXP~1E~1048.1--5937 lies on the edge of the Carina Nebula, a bright
and confused region at all wavelengths.  Jones (1973\nocite{jon73})
identified a region of non-thermal radio emission towards the
Carina Nebula, which he proposed as a candidate SNR, G287.8--0.5.
While other radio observations have failed to find evidence for such
a source (\cite{ret83}; \cite{whi94}), data at X-ray, optical and
millimeter wavelengths provide evidence for recent supernova activity 
in the Carina region (\cite{tsk91}; \cite{chu93b}; \cite{tgct98}).
Thus, while it seems likely that 1E~1048.1--5937 is in a region
of massive star formation, there is no indication of the specific event
associated with its formation.

The 7-second pulsar AX~J1845--0258 is highly time-variable, but
otherwise has properties consistent with it being an AXP (\cite{tkk+98};
\cite{gv98};\cite{vgtg00}).  It lies within $40''$ of the center of the
young SNR~G29.6+0.1, with a probability of random alignment of 
$<2\times10^{-3}$ (\cite{ggv99}).  As in the cases discussed above, 
it seems likely that the AXP and the surrounding SNR are physically
associated.

Finally, the X-ray point source RX~J1838.4--0301 lies within
the SNR~G28.8+1.5, and was once considered an AXP
candidate on the basis of a marginally significant detection
of 5.5-s pulsations in {\em ROSAT}\ data (\cite{sch94}).
However, RX~J18384.4--0301 shows significant X-ray flares, suggesting
that it likely corresponds to coronal emission from a positionally
coincident K5 star (\cite{mbn97}). Furthermore, the
pulsations reported by Schwentker (1994\nocite{sch94})
were not detected in 
subsequent {\em ASCA}\ observations (\cite{smm+00}).
Thus there currently seems little evidence that RX~J1838.4--0301 is
an AXP or is associated with G28.8+1.5.

\subsection{Overall Properties}
\label{sec_axp_overall}

In the first part of Table~\ref{tab_snrs} we summarize the properties of
the three AXPs for which there are plausible associations with SNRs. For
each AXP we list the estimated age, $t_{\rm SNR}$, and distance,
$d_{\rm SNR}$, for the SNR, the offset of the AXP from the SNR's
geometric center, $\Delta\theta$, and the angular radius of the SNR,
$\theta_{\rm SNR}$.
In all three cases the associated SNR is reasonably circular and its center
is well-defined, so that it is straightforward to estimate this offset.

The ratio of the offset to the SNR radius,
$\beta = \Delta\theta/\theta_{\rm SNR}$, is also listed
in Table~\ref{tab_snrs}. In all cases $\beta \ll 1$, indicating
that each AXP lies very close to the center of its SNR.
As argued above, these alignments imply a very small probability
that an AXP and SNR lie in the same part of the sky simply by chance.

If we assume that each AXP is associated with its coincident SNR and
was born at its center, the values of $t_{\rm SNR}$, $d_{\rm SNR}$
and $\Delta\theta$ allow one
to calculate an implied transverse velocity, $V_T$, for the AXP. In all three
cases the implied velocity is $<$500~\kms. These upper limits
and the corresponding small values of $\beta$ are both
entirely consistent with the properties of the youngest radio pulsars
and their associated SNRs (Gaensler \& Johnston 1995a,b\nocite{gj95b,gj95c};
\cite{kas96}).

The ages of their associated SNRs indicate that the three AXPs in
Table~\ref{tab_snrs} are all young objects, with ages less than $10^4$~yr.
However, the absence of SNRs around the remaining three AXPs does not
necessarily imply that these other sources are older objects.  Many SNRs
occur in low density regions and so do not produce detectable emission
(\cite{ksbg80}; \cite{gj95c}). Indeed of the eight known radio
pulsars with characteristic ages less than $10^4$~yr, four have
no associated SNR at radio wavelengths.\footnote{The four 
radio pulsars with no
SNR are the Crab Pulsar (\cite{fkcg95}), PSR~J0537--6910 (\cite{ldh+00}),
PSR~B1610--50 (\cite{sgj99}) and PSR~J1617--5055 (\cite{kcm+98}). While
the first two of these power associated synchrotron nebulae, they
show no evidence for a surrounding SNR blast-wave.} If the supernovae
which produce AXPs occur in similar environments to those which form
radio pulsars then, just as for young radio pulsars, we expect that
$\sim$50\% of young AXPs will lack SNR associations. We therefore argue that
observations are consistent with all six AXPs being young neutron stars
with ages $\la10$~kyr. 
As argued by Gaensler \etal\ (1999\nocite{ggv99}),
this implies a lower limit on the Galactic birth-rate for AXPs of one
per 1700~yr.

The young ages inferred for AXPs can place constraints on both
magnetar and accretion models for these sources.
In the context of the magnetar model, our upper limit
on AXP ages, combined with their narrow range
of spin periods, can be explained only if these
sources undergo rapid 
magnetic field decay, as can result from a Hall cascade
in the neutron star crust
(\cite{cgp00}). Such a model specifically predicts
that an AXP of true age $\sim10^4$ will have a characteristic
age $\tau = P/2\dot{P} \sim 10^5$, which indeed is
the case for AXP~1E~2259+586 ($\tau \sim 225$~kyr; \cite{kcs99})
in the SNR~CTB~109 (age $\sim$ 10~kyr; \cite{rp97}).

Any viable accretion model must produce sufficient torque to 
spin down an AXP from its presumed rapid birth period ($\ll1$~s)
to its current spin-period ($\sim10$~s) in less than $10^4$~yr.
While this rapid braking rules out many standard accretion scenarios
(\cite{vg97}), Chatterjee \etal\ (2000\nocite{chn00}) propose
a model in which an AXP is a $\sim10^{13}$~G neutron
star  which accretes from a fall-back disk of supernova debris.
In this model, the neutron star is initially in a propellor
phase in which its X-ray luminosity is too low to be detected.
After a few thousand years, the AXP will slow down sufficiently
that it can begin to accrete and will become X-ray bright.
However, the mass accretion rate will steadily decline
as the disk is depleted, and after $\sim10$~kyr the
AXP will again become too faint to be detected. This model
thus predicts that AXPs will only be observed with ages
$\sim10^4$~yr and with a narrow range
of spin periods, as is observed.

Marsden \etal\ (2001\nocite{mlrh01}) have also recently considered
associations of AXPs with SNRs. In discussing such systems, they include
the pairings \rxj\ / G346.6--0.2 and 1E~1048.1--5937
/ G287.8--0.5, both of which we have argued in Section~\ref{sec_axp_other}
above to be
spurious associations.  For the remaining three associations (listed in
Table~\ref{tab_snrs}), Marsden \etal\ (2001)
assume each SNR to be in the Sedov phase of evolution and to have
resulted from a supernova of kinetic energy $E_0 = 10^{51}$~erg. They then
use the SNR's estimated age, $t_{\rm SNR}$ and radius, $R_{\rm SNR}$,
to infer an ambient density, $n_0$.  They conclude that the supernovae
which form AXPs (and SGRs) occur in regions of significantly
higher density regions than those which make radio pulsars. 
Marsden \etal\ (2001\nocite{mlrh01}) argue that this result
favors accreting
models for AXPs, and specifically propose
that the neutron star either accretes gas
as it overtakes the slowly-expanding SNR shell, or forms
an accretion disk from material pushed back by the encounter of the SNR 
with dense material. In either case, a high ambient density 
for AXPs suggests that their properties are due to their
environment rather than are intrinsic to the 
source, a conclusion which would argue against
the magnetar hypothesis.

However, there are a number of deficiencies with this argument.  First,
Kes~73 and possibly G29.6+0.1 are very young SNRs, which may
not yet be in the Sedov phase. In this case, the calculation 
used to infer an ambient density is not valid.
Second, for SNRs in the Sedov phase the ambient density
depends on other parameters as
$n_0 \propto E_0\, t_{\rm SNR}^2\, R_{\rm SNR}^{-5}$.  Uncertainties
in $t_{\rm SNR}$ and $E_0$ of a factor of two, along with a 20\% uncertainty
in the distance (all quite reasonable for Galactic SNRs), result in
an uncertainty of two orders of magnitude in any estimate of $n_0$.

Regardless of the uncertainties in these calculations, 
the conclusion that AXPs
occur predominantly  in denser regions ($n_0 > 0.1$~cm$^{-3}$) than do
radio pulsars ($n_0 \sim 0.001$~cm$^{-3}$) can be entirely
attributed to
selection effects. 
AXPs have generally been discovered serendipitously in
X-ray observations of bright SNRs, the latter 
which are generally only detectable in 
high density regions (\cite{ksbg80}; \cite{gj95c}).  
On the other hand, young radio pulsars have
mostly been detected in all-sky surveys, and their inferred values of
$n_0$ reflect the fact that most of the interstellar medium by volume is of low
density.  When only radio pulsars associated with SNRs are considered,
ambient densities $n_0 \sim 0.2$~cm$^{-3}$ are inferred (\cite{fgw94}),
indicating that there is no obvious difference in ambient density between
SNRs associated with radio pulsars and those containing AXPs.

\subsection{Radio Emission from AXPs}

A detection of radio pulsations from an AXP would make a strong
case that these sources are isolated neutron stars rather than
accreting systems. Furthermore, this would make these sources
amenable to radio timing observations and would provide
distance estimates from their dispersion measures. 

We have shown that there is no radio emission
from \rxj\ or \4u\ down to limits of 3~mJy and 0.3~mJy respectively
(5$\sigma$ limits at 1.4 GHz). Because these limits are
determined from continuum images, they are more constraining
than comparable non-detections from pulsed searches, which
can have reduced sensitivity at long periods.
Similar non-detections of radio emission from
the AXPs 1E~2259+586 (0.08~mJy; \cite{cjl94}),
and 1E~1841--045 (0.6~mJy; \cite{kbhc85}) have
led Baring \& Harding (1998\nocite{bh98b}) to argue
that AXPs are ``radio-quiet''. If AXPs are magnetars,
this could result from photon-splitting in their 
magnetospheres, which prevents pair-production and
thus suppresses the radio pulse mechanism.

The distances to these AXPs imply upper limits
on their 1.4-GHz radio luminosities of
1.2, 2.0, 29 and 320~mJy~kpc$^2$ for
\4u, 1E~2259+586, 1E~1841--045
and \rxj\ 
respectively.\footnote{We have adopted distances to \rxj\ and \4u\
of 10 and 2~kpc respectively, and have omitted
the factor of $4\pi$ as is usual for pulsar
radio luminosities.} Lyne \etal\ (1998\nocite{lml+98})
derive a 400-MHz luminosity function for potentially
observable
radio pulsars (i.e.\ all pulsars in the Galaxy
beaming towards us, whether detected by current searches
or as yet undiscovered). 
Scaling this luminosity distribution
to an observing frequency of 1.4~GHz by assuming a typical
pulsar spectral index of $\alpha = -2$, we find that 
$>$60\% of potentially observable radio pulsars 
have pulsed luminosities below 1~mJy~kpc$^2$,
fainter than the deepest limits obtained towards the AXPs. 
The fact that 
that this small set of X-ray selected sources has not been
detected at radio wavelengths therefore does not place any strong
constraints on their intrinsic radio properties.

Furthermore, models
for pulsar beaming generally predict that slower-spinning
pulsars have narrower radio beams. Indeed
the slowest radio pulsar, PSR~J2144--3933 
(for which $P=8.5$~s, comparable to that of the AXPs),
has the narrowest known pulse, of width less than $1^\circ$ 
of the pulse phase (\cite{ymj99}).
If we assume a population of radio pulsars in which the
magnetic axis is randomly oriented with respect both  to
the pulsar spin axis and to the line-of-sight, 
\footnote{In magnetar models for the AXPs,
the X-ray emission originates from the neutron star surface (\cite{hh97})
and thus suffers significant gravitational bending.
The X-ray beams will thus be very broad, and the 
fact that we observe X-ray pulsations from the AXPs
is not inconsistent with the assumption of randomly oriented
radio beams.}
then for $P=10$~s (appropriate for the AXPs),
the fraction of pulsars which would be beamed towards us is
estimated to be 3--5\% (e.g.\ \cite{big90b};
\cite{tm98}).
Thus even if all the AXPs are radio pulsars,
it is likely that none of them are beamed in our direction.

\section{Associations of SGRs with SNRs}
\label{sec_sgrs}

Of the four (possibly five) SGRs identified to date,
in all cases associations
have been claimed with nearby SNRs (\cite{hur00b}). Below we review
recent results on each system, and use similar criteria
as for the AXPs to assess the validity of the proposed SNR associations.

\subsection{SGR~0526--66 and SNR~N49}

SGR~0526--66 was discovered as a result of its intense $\gamma$-ray
activity on 5~Mar~1979, data which
also contained an 8-second periodicity (\cite{bch+79};
\cite{tekl80}).  This $\gamma$-ray
emission was localized to a small
0.1~arcmin$^2$ region of the Large Magellanic Cloud (LMC) which overlaps
the SNR~N49 (\cite{cdt+82}). The X-ray point source RX~J05260.3--660433
falls within this error box and lies on the rim of the SNR (\cite{rkl94};
\cite{mrlp96});
this source is presumed to be the X-ray counterpart
to the SGR.

The probability of a chance alignment between SGR~0526--66 and SNR~N49 was
estimated by Felten (1982\nocite{fel82}) to be $\sim10^{-3}$. There have
been many new results on SNRs in the LMC since this estimate was made.
We consequently here make a revised calculation as to the probability
of coincidence between the SNR and SGR.

The most recent catalog of LMC SNRs is that of Williams \etal\
(1999\nocite{wcd+99}),
who list 37 confirmed SNRs. The total solid angle subtended by these
SNRs is $\sim180$~arcmin$^2$.  If we assume the LMC to be delineated
by a circle of radius $3^\circ$ and that SNRs are randomly
distributed throughout this region, then using the same approach as in
Section~\ref{sec_axps_rxj}
above the probability of the SGR falling on the rim of an unrelated
SNR is 0.2\%, double the estimate of Felten (1982\nocite{fel82}).

However, because both SNRs and SGRs are believed to be formed in supernova
explosions, their distributions should be similar to those of massive
stars and star-forming regions within the LMC, neither of which are
uniformly distributed throughout the entire galaxy.  The above calculation
therefore considers too large a possible area for these objects, and is
an underestimate of the probability of a spurious association.

\begin{figure}[htb]
\vspace{3cm}
\centerline{\psfig{file=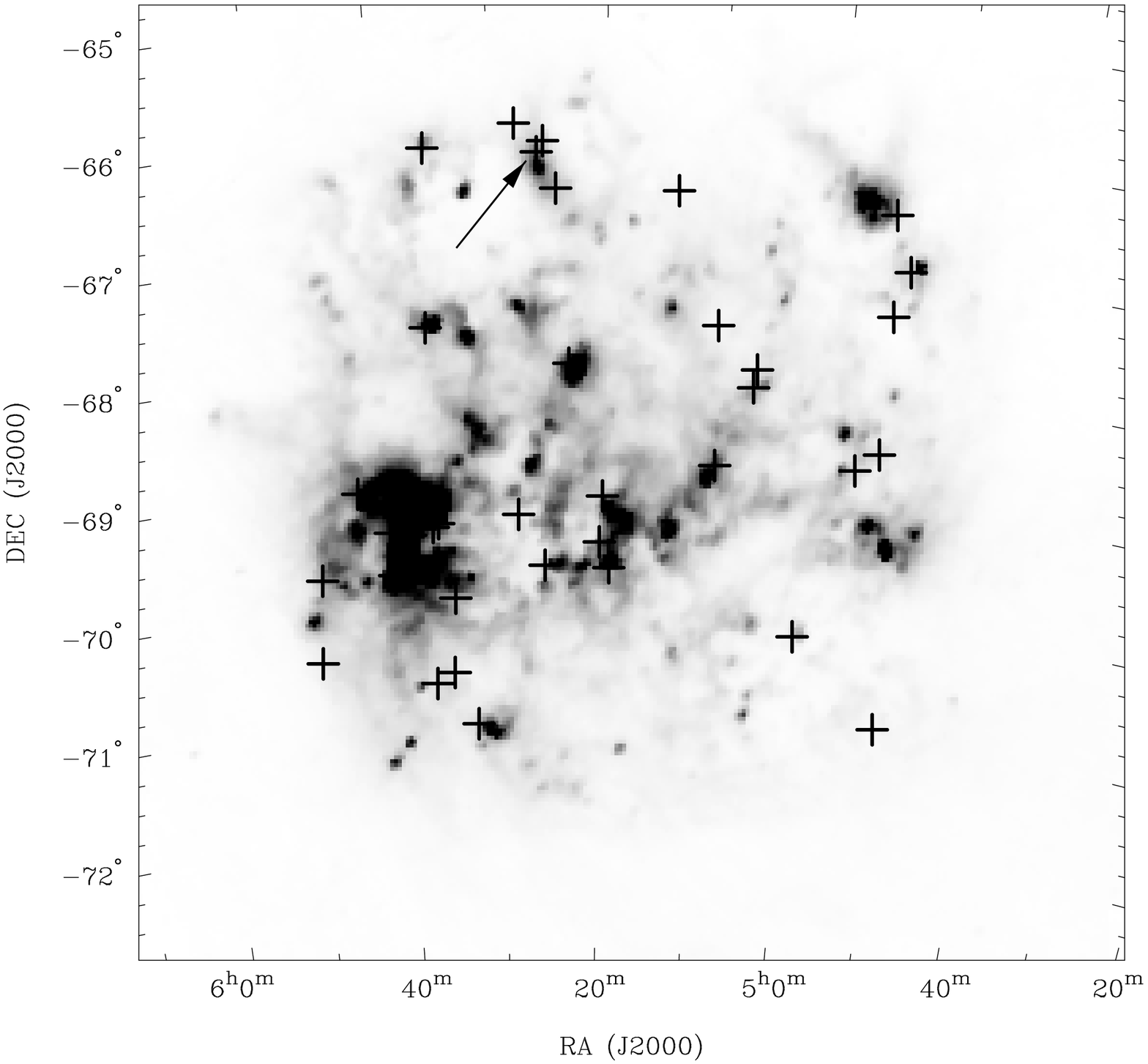,width=7in}}
\caption{{\em IRAS}\ 60~$\mu$m image of the Large Magellanic
Cloud, with a greyscale range of 0 to 50 MJy~sr$^{-1}$.
The positions of SNRs from the catalog
of Williams \etal\ (1999\protect\nocite{wcd+99}) are marked by ``+'' symbols;
the arrow points to SNR~N49. It can be seen that SNRs are not randomly
distributed throughout the LMC, but trace the spiral structure seen
in the infrared.}
\label{fig_lmc}
\end{figure}

In Figure~\ref{fig_lmc} we show an {\em IRAS}\ 60~$\mu$m image of the LMC,
on which the positions of SNRs from the catalog of Williams \etal\
(1999\nocite{wcd+99})
are marked. It is clear that the distribution of SNRs within the
LMC is indeed far from uniform, and follows the
multi-armed spiral pattern traced in the infra-red. If
we only consider those parts of
the LMC in which there is bright infrared
emission ($\Sigma_{60\, {\rm \mu m}} >
10$~MJy~sr$^{-1}$), the total area
under consideration is then $\sim7$~deg$^2$,
and the probability of a chance coincidence between
SGR~0526--66 and a SNR rises to 0.7\%.
Alternatively, the scale length for the distribution of
OB~stars in the LMC is 1.6~kpc (\cite{wn01}),
corresponding to an area of $\sim10$~deg$^2$ (for
a distance of 50~kpc) and thus a probability of random
alignment of 0.5\%.

There are certainly many LMC SNRs still to be discovered, especially
given the difficulty in identifying SNRs in complex star-forming regions
(\cite{ck88}). The above calculations
are thus likely to be underestimates, and the true probability of
coincidental alignment may be as large as several percent. Furthermore,
as we have discussed above for the case of \rxj, a low probability of
chance alignment does not imply that the SGR and SNR were formed
in the same supernova explosion, but possibly that two separate supernovae
occurred near each other. The many examples
of closely-grouped SNRs in the LMC indeed demonstrate the
clustered nature of their progenitors (\cite{clg+93}; \cite{scm+94}; 
\cite{wcd+97}). 

These considerations suggest that the claimed association between
SGR~0526--66 and N49, while by far the most likely of the SGR/SNR
associations (see further discussion below), is considerably less
convincing than the cases for AXPs in SNRs considered above.
Doubt has also recently been cast on this association by Kaplan \etal\
(2001\nocite{kkv+01}), who infer an age for the SGR of $\sim$1000~yr 
on the basis of its energetics and broad-band spectrum. This is clearly
inconsistent with the age of 5--16~kyr estimated for the SNR
(\cite{shu83}; \cite{vblr92}).

\subsection{SGR~1806--20 and G10.0--0.3}

SGR~1806--20 was originally localized to a small region centered on the
unusual SNR~G10.0--0.3 (\cite{abh+87};
\cite{mtk+94}). However, a re-analysis of $\gamma$-ray
data for SGR~1806--20 demonstrates its most likely position to be
offset by $15''$ from the center of G10.0--0.3 (\cite{hkc+99}).
Recent observations
of the region using the
{\em Chandra X-ray Observatory}\ have confirmed this refined
position (\cite{kap01}). 

Meanwhile, VLA observations of G10.0--0.3 have shown a
centrally-condensed, changing morphology and an unusually steep spectrum,
with no evidence for a blast-wave and/or a
supernova explosion (\cite{kfk+94}; \cite{vfk95}; \cite{fvk97}).
The classification of G10.0--0.3 as a SNR 
therefore seems to have been erroneous; 
it rather seems to be powered by some central
object unrelated to SGR~1806--20, possibly the massive star
identified by van Kerkwijk \etal\ (1995\nocite{vkmn95}).

We thus conclude that there is no SNR associated with SGR~1806--20.

\subsection{SGR~1900+14 and G42.8+0.6}

SGR~1900+14 has been identified with an unresolved
X-ray source located $5'$ outside the rim of
the SNR~G42.8+0.6 (\cite{hlk+99}). As is
the case for \rxj, this source falls in a complicated
region in the inner Galaxy. Considering the region
$35^\circ < l < 45^\circ$, $|b| < 1^\circ$, 
we find by the same approach as used earlier
that there is a $\sim$4\%  probability that
the proximity to a SNR is simply by chance. As discussed above,
this low probability may simply represent
the natural spatial clustering of supernova explosions rather
than any specific association between the SGR and SNR.
Furthermore, a radio pulsar, PSR~J1907+0918, has recently been identified
just $2'$ from SGR~1900+14, and has a characteristic
age of only 38~kyr (\cite{lx00}). The case
for an association between the radio pulsar and SNR~G32.8+0.6
is far from compelling, but is just as plausible as the SGR/SNR
association, further weakening the argument for the latter.

Finally, we note that Vrba \etal\ (2000\nocite{vhl+00}) have
discovered a cluster of massive stars separated from SGR~1900+14
by just a few arcsec. They propose an association between the SGR and the
star cluster, which would immediately rule out an association 
with SNR~G42.8+0.6.

We conclude that there is a significant likelihood
of random alignment between SGR~1900+14 and
G42.8+0.6, and that there is little convincing
evidence in favor of a physical association given the
complex nature of this part of the sky.

\subsection{SGR~1627--41 and G337.0--0.1}

SGR~1627--41 has an X-ray counterpart which shows it
to be embedded in the complicated radio region CTB~33 (\cite{wkv+99}).
CTB~33 was originally classified as an SNR/\HII\ complex (\cite{sg70c}),
an interpretation confirmed by recent high-resolution observations
which show this source to be a large \HII\ 
region in which a compact SNR, G337.0--0.1, is
embedded (\cite{sggf97}). The X-ray localization
for SGR~1627--41 puts it $\sim30''$ outside
the rim of the SNR (\cite{hsk+00}). Smith \etal (1999\nocite{sbl99})
estimate the probability of a random alignment between the SNR
and SGR to be $\sim$5\%.

A common distance of 11.0~kpc has been determined independently for
SNR~G337.0--0.1 (\cite{fgr+96}) and SGR~1627--41 (\cite{ccdd99}).
However, this simply indicates
that the SNR and SGR are probably part of the same star-forming
complex which includes CTB~33; it does not make a strong case
that they correspond to the same supernova explosion.

\subsection{SGR~1801--23}

A possible fifth SGR, 1801--23, has been reported by Cline \etal\ 
(2000\nocite{cfg+00}). The error box for this source is
very elongated, extending across almost four degrees of
the Galactic Plane in a complex region near the Galactic
Center. Although this error circle passes through or near
$\sim$7 SNRs, any such error circle drawn randomly on
the sky in this region would do so. No serious case
can be made for an association with any particular SNR until
this SGR is confirmed and its position refined.

\subsection{Overall Properties}

In the second half of Table~\ref{tab_snrs}, we have listed the three
SGRs which lie near SNRs. As
in Section~\ref{sec_axp_overall} 
for the AXPs, for each system we have listed the
age, distance and radius of the SNR, the offset of
the SGR with respect to the SNR's center, the normalized
offset, $\beta$, and the implied transverse
velocity, $V_T$. In stark distinction to the AXPs,
for all three SGRs we find that $\beta \ge 1$. 

No estimate of an age or distance is available for G42.8+0.6,\footnote{The
$\Sigma-D$\ relation is not a valid method of determining distances to
individual SNRs (e.g.\ \cite{gre84}), and distances and ages which have been
quoted for this SNR using this method are totally unreliable.} but for
SNRs~N49 and G337.0--0.1, transverse velocities which are at the
upper end of,
or even beyond, the velocity distribution of the radio pulsar population 
(\cite{lbh97}; \cite{cc98})
are inferred. These velocities appear inconsistent
with the low transverse velocities inferred for the AXPs.

The small statistics available thus suggest that if SGRs
are associated with their nearby SNRs, then they must be
a higher velocity population
than are AXPs or radio pulsars. 
Indeed it has been proposed that magnetars should 
have unusually high space velocities ($\ga1000$~\kms)
as a result of anisotropic neutrino emission immediately
after core-collapse (\cite{dt92a}; \cite{td93a}). 

However, these associations with SNRs
rule out any relationship or evolutionary connection with the AXPs.
If SGRs are a younger population than the AXPs, as has been proposed by 
Kouveliotou \etal\ (1998\nocite{kds+98}),
then SGRs should generally have smaller values of $\beta$ than do AXPs,
which is clearly not the case. On the other hand, if SGRs are an
older incarnation of the AXPs (\cite{ggv99}; \cite{gvd99}), then the 
SGRs should have values of $\beta$ larger than for
the AXPs (as is observed), but the velocity
distributions
of the two populations should
be similar.\footnote{While the ``electromagnetic rocket'' effect 
proposed by Harrison \&
Tademaru (1975\nocite{ht75}) can cause a neutron star's space velocity
to steadily increase, this effect is negligible at the long periods of
AXPs and SGRs (\cite{lcc01}).}
Thus if all the associations in Table~\ref{tab_snrs} are genuine,
then AXPs and SGRs must represent two discrete populations of
object. This could result either if SGRs
are magnetars  and AXPs are accreting systems, or if
there are two types of magnetar --- a population
with extreme ($\sim10^{15}$ G) magnetic fields and 
high velocities (the SGRs), and 
a separate group with lower magnetic fields ($\sim10^{14}$~G)
and lower velocities (the AXPs).

Disregarding their values of $\beta$ and $V_T$,
AXPs and SGRs have remarkably similar
properties (e.g.\ \cite{kkm+00}; \cite{kgc+01}). There is 
thus considerable reluctance to conclude that the
two populations are not related in some way.  The only
way this AXP/SGR connection
can be maintained is if one abandons the associations between SGRs and
SNRs, and therefore removes the discrepancy in the velocity distributions
of the AXP and SGR populations. 
As argued earlier, none of the SGR/SNR associations are particularly
compelling. It is worth noting that of the claimed
associations between radio pulsars and SNRs for which $\beta \ge 1$,
almost all have been subsequently argued to be spurious (e.g.\
\cite{gj95c}; \cite{njk96}; \cite{kcm+98}; \cite{sgj99}).

If SGRs are neutron stars, but have ages of 50--100~kyr and
have low space velocities, then we would expect their associated SNRs to
have faded, but for these sources to still be near regions of supernova
and star-forming activity, as is observed. In this case the
the data are consistent with the hypothesis that SGRs
and AXPs are related sources, but imply that the SGRs represent
an older or longer-lived population whose SNRs have dissipated.

Given this result, 
it is tempting to argue that AXPs evolve into SGRs.
However, this scenario is problematic in that
if AXPs are magnetars, then their youth and narrow range
of spin-periods argue that their magnetic fields decay
rapidly to strengths below $\sim10^{13}$~G (\cite{cgp00}).
It is then hard to see how AXPs can evolve into SGRs, the
latter of which are believed to have
magnetic field strengths $\ga10^{15}$~G. Furthermore,
the AXPs and SGRs
have similar period distributions: 5--8~sec for SGRs, and 6--12~sec
for AXPs.  If SGRs are 5--10 times older than AXPs as proposed here,
and if one extrapolates the steady spin-down seen in several AXPs 
(\cite{gvd99}; \cite{kcs99}) to such ages, we would then
expect the SGRs to have periods $\gg$10~s, which is not observed. 
We note that
some AXPs and SGRs do not show smooth spin-down (\cite{kgc+01};
\cite{wkv+99b}), while there is evidence in the case
of 1E~2259+586 for a short period of spin-{\em up} (\cite{bs96}).
It has indeed been argued that magnetars can undergo
periods of very low spin-down or even spin-up depending
on the level of internal activity
(\cite{tdw+00}). These effects must be significant
if AXPs are to evolve into SGRs.

\section{Conclusions}

We have presented radio observations of two Anomalous X-ray Pulsars,
\rxj\ and \4u. \rxj\ lies in a complex region which includes
the candidate supernova remnant~\rxjsnr, but is not likely to be 
associated with this or any other SNR. No extended radio emission is
seen in the vicinity of \4u; even the faintest known SNR would have been
detected.

The three associations of AXPs with SNRs provide strong evidence
for a physical connection between the two types of object.
These associations imply that AXPs are young
($<$10~kyr) neutron stars, with low space velocities ($<$500~\kms).
This result is consistent with AXPs being ``magnetars''
which undergo rapid field decay, but also with
models in which AXPs are lower-field neutron stars
accreting from a supernova fallback disk.
We argue that there is no evidence that the SNRs associated with
AXPs expand into a denser environment than do the SNRs associated
with radio pulsars, contrary to the ``nurture'' model for
AXPs proposed by Marsden \etal\ (2001\nocite{mlrh01}).
Despite the fact that no AXP has been detected at radio
wavelengths, these data provide no evidence that
AXPs are radio-quiet. Even if all AXPs are radio pulsars,
it is highly likely that they are either all beaming away from us
or are too faint to detect.

The Soft Gamma-Ray Repeaters have been proposed to all be
associated with SNRs.  We review the associations claimed
between SGRs and SNRs --- in one case the classification
as a SNR was erroneous,
while in all other cases the SGR is on the edge of
or outside the nearest SNR. We find a significant
probability that the claimed SGR/SNR associations
are spurious, implying that the SGRs
are an older or longer-lived population than are the AXPs.
If the SGR/SNR associations
are valid, they imply that SGRs are high
velocity ($\ga1000$~\kms) objects. In this case, the data
are inconsistent with there being any
link between the SGR and AXP populations. In either case,
the fact that the SGRs all lie near SNRs, \HII\ regions and
massive star clusters is still consistent with the hypothesis that they are
relatively young neutron stars.

The high velocities suggested for the SGRs from their claimed
SNR associations can be tested
by proper motion measurements with {\em Chandra}\
over the next few years. These results will provide significant new
input into the debate over SGR/SNR associations, and on the relationship
between SGRs and AXPs.

\begin{acknowledgements}

We thank Barry Clark for generously rescheduling
a failed observing run, Fred Seward and Kate Brooks for information
on SNRs in the Carina region, and
Bob Rutledge for discussions on probability.
The National Radio Astronomy Observatory is a facility of the National
Science Foundation operated under cooperative agreement by Associated
Universities, Inc.  This paper has used data obtained as a Guest User
of the  Canadian Astronomy Data Center, and has made
use of the
NASA Astrophysics Data System, the CDS SIMBAD database and
NASA's SkyView facility. Some of this
work was completed while B.M.G. was at the Institute
for Theoretical Physics, which is supported
by NSF under Grant No. PHY99--07949. This research was
also supported by NASA through
Hubble Fellowship grant HF-01107.01-98A (B.M.G.),
contract NAS8-39073 and grant NAG5-4803 (P.O.S.),
and LTSA grant NAG5--22250 (E.V.G. and G.V.). 

\end{acknowledgements}


\bibliographystyle{apj1}
\bibliography{journals,modrefs,psrrefs,crossrefs}

\begin{thebibliography}{}

\bibitem[{Atteia} \etal  1987]{abh+87}
{Atteia}, J.-L. \etal  1987, { ApJ}, {\rm 320}, L105.

\bibitem[Barat \etal  1979]{bch+79}
Barat, C., Chambon, G., Hurley, K., Niel, M., Vedrenne, G., Estulin, I.~V.,
  Kurt, V.~G., \& Zenchenko, V.~M. 1979, { A\&A}, {\rm 79}, L24.

\bibitem[Baring \& Harding 1998]{bh98b}
Baring, M.~G. \& Harding, A.~K. 1998, { ApJ}, {\rm 507}, L55.

\bibitem[Baykal \& Swank 1996]{bs96}
Baykal, A. \& Swank, J. 1996, { ApJ}, {\rm 460}, 470.

\bibitem[Biggs 1990]{big90b}
Biggs, J.~D. 1990, { MNRAS}, {\rm 245}, 514.

\bibitem[Braun, Goss, \& Lyne 1989]{bgl89}
Braun, R., Goss, W.~M., \& Lyne, A.~G. 1989, { ApJ}, {\rm 340}, 355.

\bibitem[Cao \etal  1997]{ctpb97}
Cao, Y., Terebey, S., Prince, T.~A., \& Beichman, C.~A. 1997, { ApJS}, {\rm
  111}, 387.

\bibitem[Caraveo 1993]{car93}
Caraveo, P.~A. 1993, { ApJ}, {\rm 415}, L111.

\bibitem[Caswell \& Haynes 1987]{ch87b}
Caswell, J.~L. \& Haynes, R.~F. 1987, { A\&A}, {\rm 171}, 261.

\bibitem[Caswell \etal  1995]{cvew+95}
Caswell, J.~L., Vaile, R.~A., Ellingsen, S.~P., Whiteoak, J.~B., \& Norris,
  R.~P. 1995, { MNRAS}, {\rm 272}, 96.

\bibitem[Chatterjee \& Hernquist 2000]{ch00}
Chatterjee, P. \& Hernquist, L. 2000, { ApJ}, {\rm 543}, 368.

\bibitem[{Chatterjee}, {Hernquist}, \& {Narayan} 2000]{chn00}
{Chatterjee}, P., {Hernquist}, L., \& {Narayan}, R. 2000, { ApJ}, {\rm 534},
  373.

\bibitem[{Chu} \& {Kennicutt} 1988]{ck88}
{Chu}, Y. \& {Kennicutt}, R.~C. 1988, { AJ}, {\rm 96}, 1874.

\bibitem[Chu 1993]{chu93b}
Chu, Y.-H. 1993, { BAAS}, {\rm 182}, 27.06.

\bibitem[Chu \etal  1993]{clg+93}
Chu, Y.-H., MacLow, M.-M., Garcia-Segura, G., Wakker, B., \& Kennicutt~Jr.,
  R.~C. 1993, { ApJ}, {\rm 414}, 213.

\bibitem[Cline \etal  2000]{cfg+00}
Cline, T., Frederiks, D.~D., Golenetskii, S., Hurley, K., Kouveliotou, C.,
  Mazets, E., \& van Paradijs, J. 2000, { ApJ}, {\rm 531}, 407.

\bibitem[{Cline} \etal  1982]{cdt+82}
{Cline}, T.~L. \etal  1982, { ApJ}, {\rm 255}, L45.

\bibitem[Coe, Jones, \& Lehto 1994]{cjl94}
Coe, M.~J., Jones, L.~R., \& Lehto, H. 1994, { MNRAS}, {\rm 270}, 178.

\bibitem[Colpi, Geppert, \& Page 2000]{cgp00}
Colpi, M., Geppert, U., \& Page, D. 2000, { ApJ}, {\rm 529}, L29.

\bibitem[Corbel \etal  1999]{ccdd99}
Corbel, S., Chapuis, S., Dame, T.~M., \& Durouchoux, P. 1999, { ApJ}, {\rm
  526}, L29.

\bibitem[Cordes \& Chernoff 1998]{cc98}
Cordes, J.~M. \& Chernoff, D.~F. 1998, { ApJ}, {\rm 505}, 315.

\bibitem[Crawford \etal  2001]{cgk+00}
Crawford, F., Gaensler, B.~M., Kaspi, V.~M., Manchester, R.~N., Camilo, F., G.,
  L.~A., \& Pivovaroff, M.~J. 2001, { ApJ}, {\rm 554}.
\newblock in press (astro-ph/0012287).

\bibitem[Dar \& De~R\'{u}jula 2000]{dd00}
Dar, A. \& De~R\'{u}jula, A. 2000, {\rm }.
\newblock astro-ph/0002014.

\bibitem[Dubner \etal  1993]{dmgw93}
Dubner, G.~M., Moffett, D.~A., Goss, W.~M., \& Winkler, P.~F. 1993, { AJ}, {\rm
  105}, 2251.

\bibitem[Duncan 2000]{dun00}
Duncan, R.~C. 2000, in { Fifth Hunstville Gamma-Ray Burst Symposium, AIP
  Conference Proceedings No. 526}, ed.\ R.~M. Kippen, R.~S. Mallozzi, \& G.~J.
  Fishman, (New York: American Institute of Physics), (astro-ph/0002442).

\bibitem[Duncan \& Thompson 1992]{dt92a}
Duncan, R.~C. \& Thompson, C. 1992, { ApJ}, {\rm 392}, L9.

\bibitem[Ellison, Slane, \& Gaensler 2001]{esg01}
Ellison, D.~C., Slane, P., \& Gaensler, B.~M. 2001, { ApJ}, {\rm }.
\newblock submitted.

\bibitem[{English} \etal  1998]{eti+98}
{English}, J. \etal  1998, { Publ. Astron. Soc. Austral.}, {\rm 15}, 56.

\bibitem[Fahlman \& Gregory 1981]{fg81}
Fahlman, G.~G. \& Gregory, P.~C. 1981, { Nature}, {\rm 293}, 202.

\bibitem[Felten 1982]{fel82}
Felten, J.~E. 1982, in { Proceedings of the 17th International Cosmic Ray
  Conference}, (Gif-sur-Yvette, Essonne, France: Commissariat a l'Energie
  Atomique), p.~52.

\bibitem[Frail \etal  1996]{fgr+96}
Frail, D.~A., Goss, W.~M., Reynoso, E.~M., Giacani, E.~B., Green, A.~J., \&
  Otrupcek, R. 1996, { AJ}, {\rm 111}, 1651.

\bibitem[Frail, Goss, \& Whiteoak 1994]{fgw94}
Frail, D.~A., Goss, W.~M., \& Whiteoak, J. B.~Z. 1994, { ApJ}, {\rm 437}, 781.

\bibitem[Frail \etal  1995]{fkcg95}
Frail, D.~A., Kassim, N.~E., Cornwell, T.~J., \& Goss, W.~M. 1995, { ApJ}, {\rm
  454}, L129.

\bibitem[Frail, Kassim, \& Weiler 1994]{fkw94}
Frail, D.~A., Kassim, N.~E., \& Weiler, K.~W. 1994, { AJ}, {\rm 107}, 1120.

\bibitem[Frail, Vasisht, \& Kulkarni 1997]{fvk97}
Frail, D.~A., Vasisht, G., \& Kulkarni, S.~R. 1997, { ApJ}, {\rm 480}, L129.

\bibitem[Gaensler 2000]{gae00}
Gaensler, B.~M. 2000, in { Pulsar Astronomy --- 2000 and Beyond, {IAU}
  Colloquium 177}, ed.\ M. Kramer, N. Wex, \& R. Wielebinski, (San Francisco:
  Astronomical Society of the Pacific), 703.

\bibitem[Gaensler, Gotthelf, \& Vasisht 1999]{ggv99}
Gaensler, B.~M., Gotthelf, E.~V., \& Vasisht, G. 1999, { ApJ}, {\rm 526}, L37.

\bibitem[Gaensler \& Johnston 1995a]{gj95b}
Gaensler, B.~M. \& Johnston, S. 1995a, { Publ. Astron. Soc. Austral.}, {\rm
  12}, 76.

\bibitem[Gaensler \& Johnston 1995b]{gj95c}
Gaensler, B.~M. \& Johnston, S. 1995b, { MNRAS}, {\rm 277}, 1243.

\bibitem[Gavaramadze 2001]{gav01}
Gavaramadze, V.~V. 2001, { MNRAS}, {\rm }.
\newblock submitted (astro-ph/0005572).

\bibitem[{Gotthelf} \& {Vasisht} 1998]{gv98}
{Gotthelf}, E.~V. \& {Vasisht}, G. 1998, { New Astronomy}, {\rm 3}, 293.

\bibitem[{Gotthelf}, {Vasisht}, \& Dotani 1999]{gvd99}
{Gotthelf}, E.~V., {Vasisht}, G., \& Dotani, T. 1999, { ApJ}, {\rm 522}, L49.

\bibitem[Green \etal  1999]{gcl98}
Green, A.~J., Cram, L.~E., Large, M.~I., \& Ye, T. 1999, { ApJS}, {\rm 122},
  207.
\newblock (http://www.astrop.physics.usyd.edu.au/MGPS/).

\bibitem[Green 1984]{gre84}
Green, D.~A. 1984, { MNRAS}, {\rm 209}, 449.

\bibitem[Green 1989]{gre89}
Green, D.~A. 1989, { MNRAS}, {\rm 238}, 737.

\bibitem[Green 2000]{gre00}
Green, D.~A. 2000, { A {C}atalogue of {G}alactic {S}upernova {R}emnants (2000
  {A}ugust {V}ersion)}, (Cambridge: Mullard Radio Astronomy Observatory).
\newblock (http://www.mrao.cam.ac.uk/surveys/snrs/).

\bibitem[Harding, Contopoulos, \& Kazanas 1999]{hck99}
Harding, A.~K., Contopoulos, I., \& Kazanas, D. 1999, { ApJ}, {\rm 525}, L125.

\bibitem[Harrison \& Tademaru 1975]{ht75}
Harrison, E.~R. \& Tademaru, E. 1975, { ApJ}, {\rm 201}, 447.

\bibitem[Helfand \etal  1989]{hvbl89}
Helfand, D.~J., Velusamy, T., Becker, R.~H., \& Lockman, F.~J. 1989, { ApJ},
  {\rm 341}, 151.

\bibitem[{Heyl} \& {Hernquist} 1997]{hh97}
{Heyl}, J.~S. \& {Hernquist}, L. 1997, { ApJ}, {\rm 489}, L67.

\bibitem[Hulleman, van Kerkwijk, \& Kulkarni 2000]{hvk00}
Hulleman, F., van Kerkwijk, M.~H., \& Kulkarni, S.~R. 2000, { Nature}, {\rm
  408}, 689.

\bibitem[Hurley 2000]{hur00b}
Hurley, K. 2000, in { Fifth Hunstville Symposium on Gamma Ray Bursts}, ed.\
  R.~M. Kippen, R.~S. Mallozzi, \& G.~J. Fishman, (New York: AIP Press), 763.

\bibitem[Hurley \etal  1999]{hkc+99}
Hurley, K., Kouvelioutou, C., Cline, T., Mazets, E., Golenetskii, S.,
  Frederiks, D.~D., \& van Paradijs, J. 1999, { ApJ}, {\rm 523}, L37.

\bibitem[{Hurley} \etal  1999]{hlk+99}
{Hurley}, K. \etal  1999, { ApJ}, {\rm 510}, L111.

\bibitem[{Hurley} \etal  2000]{hsk+00}
{Hurley}, K. \etal  2000, { ApJ}, {\rm 528}, L21.

\bibitem[{Israel} \etal  1999]{ics+99}
{Israel}, G.~L., {Covino}, S., {Stella}, L., {Campana}, S., {Haberl}, F., \&
  {Mereghetti}, S. 1999, { ApJ}, {\rm 518}, L107.

\bibitem[Jones 1973]{jon73}
Jones, B.~B. 1973, { Aust. J. Phys.}, {\rm 26}, 545.

\bibitem[Kafatos \etal  1980]{ksbg80}
Kafatos, M., Sofia, S., Bruhweiler, F., \& Gull, S. 1980, { ApJ}, {\rm 242},
  294.

\bibitem[Kaplan 2001]{kap01}
Kaplan, D.~L. 2001, { Mem. Soc. Astron. It.}, {\rm }.
\newblock in press.

\bibitem[Kaplan \etal  2001]{kkv+01}
Kaplan, D.~L., Kulkarni, S.~R., van Kerkwijk, M.~H., Rothschild, R.~E.,
  Lingenfelter, R.~L., Marsden, D., Danner, R., \& Murakami, T. 2001, { ApJ},
  {\rm }.
\newblock in press (astro-ph/0103179).

\bibitem[Kaspi 1996]{kas96}
Kaspi, V.~M. 1996, in { Pulsars: Problems and Progress, {IAU} Colloquium 160},
  ed.\ S. Johnston, M.~A. Walker, \& M. Bailes, (San Francisco: Astronomical
  Society of the Pacific), p.~375.

\bibitem[Kaspi, Chakrabarty, \& Steinberger 1999]{kcs99}
Kaspi, V.~M., Chakrabarty, D., \& Steinberger, J. 1999, { ApJ}, {\rm 525}, L33.

\bibitem[Kaspi \etal  1998]{kcm+98}
Kaspi, V.~M., Crawford, F., Manchester, R.~N., Lyne, A.~G., Camilo, F.,
  D'Amico, N., \& Gaensler, B.~M. 1998, { ApJ}, {\rm 503}, L161.

\bibitem[Kaspi \etal  2001]{kgc+01}
Kaspi, V.~M., Gavriil, F.~P., Chakrabarty, D., Lackey, J.~R., \& Muno, M.~P.
  2001, { ApJ}, {\rm }.
\newblock submitted (astro-ph/0011368).

\bibitem[Kassim \& Weiler 1990]{kw90}
Kassim, N.~E. \& Weiler, K.~W. 1990, { Nature}, {\rm 343}, 146.

\bibitem[Katz-Stone \& Rudnick 1997]{kr97}
Katz-Stone, D.~M. \& Rudnick, L. 1997, { ApJ}, {\rm 488}, 146.

\bibitem[Koralesky \etal  1998]{kfg+98}
Koralesky, B., Frail, D.~A., Goss, W.~M., Claussen, M.~J., \& Green, A.~J.
  1998, { AJ}, {\rm 116}, 1323.

\bibitem[Kouveliotou \etal  1998]{kds+98}
Kouveliotou, C. \etal  1998, { Nature}, {\rm 393}, 235.

\bibitem[Kriss \etal  1985]{kbhc85}
Kriss, G.~A., Becker, R.~H., Helfand, D.~J., \& Canizares, C.~R. 1985, { ApJ},
  {\rm 288}, 703.

\bibitem[Kulkarni \& Frail 1993]{kf93}
Kulkarni, S.~R. \& Frail, D.~A. 1993, { Nature}, {\rm 365}, 33.

\bibitem[Kulkarni \etal  1994]{kfk+94}
Kulkarni, S.~R., Frail, D.~A., Kassim, N.~E., Murakami, T., \& Vasisht, G.
  1994, { Nature}, {\rm 368}, 129.

\bibitem[Kulkarni \etal  2001]{kkm+00}
Kulkarni, S.~R., Kaplan, D.~L., Marshall, H.~L., Frail, D.~A., Murakami, T., \&
  Yonetoku, D. 2001, { Nature}, {\rm }.
\newblock submitted.

\bibitem[Lai, Chernoff, \& Cordes 2001]{lcc01}
Lai, D., Chernoff, D.~F., \& Cordes, J.~M. 2001, { ApJ}, {\rm 549}, 1111.

\bibitem[Lazendic \etal  2000]{ldh+00}
Lazendic, J.~S., Dickel, J.~R., Haynes, R.~F., Jones, P.~A., \& White, G.~L.
  2000, { ApJ}, {\rm 540}, 808.

\bibitem[Lorimer, Bailes, \& Harrison 1997]{lbh97}
Lorimer, D.~R., Bailes, M., \& Harrison, P.~A. 1997, { MNRAS}, {\rm 289}, 592.

\bibitem[Lorimer \& Xilouris 2000]{lx00}
Lorimer, D.~R. \& Xilouris, K.~M. 2000, { ApJ}, {\rm 545}, 385.

\bibitem[Lyne \etal  1998]{lml+98}
Lyne, A.~G. \etal  1998, { MNRAS}, {\rm 295}, 743.

\bibitem[Marsden \etal  1996]{mrlp96}
Marsden, D., Rothschild, R.~E., Lingenfelter, R.~E., \& Puetter, R.~C. 1996, {
  ApJ}, {\rm 470}, 513.

\bibitem[Marsden \etal  2001]{mlrh01}
Marsden, R., Lingenfelter, R.~E., Rothschild, R.~E., \& Higdon, J.~C. 2001, {
  ApJ}, {\rm 550}, 397.

\bibitem[McAdam, Osborne, \& Parkinson 1993]{mop93}
McAdam, W.~B., Osborne, J.~L., \& Parkinson, M.~L. 1993, { Nature}, {\rm 361},
  516.

\bibitem[Mereghetti 2000]{mer00}
Mereghetti, S. 2000, in { The Neutron Star - Black Hole Connection}, ed.\ V.
  Connaughton, C. Kouveliotou, J. {van Paradijs}, \& J. Ventura, NATO Advanced
  Study Institute, (astro-ph/9911252).

\bibitem[Mereghetti, Belloni, \& Nasuti 1997]{mbn97}
Mereghetti, S., Belloni, T., \& Nasuti, F.~P. 1997, { A\&A}, {\rm 321}, 835.

\bibitem[Mereghetti \& Stella 1995]{ms95}
Mereghetti, S. \& Stella, L. 1995, { ApJ}, {\rm 442}, L17.

\bibitem[Murakami \etal  1994]{mtk+94}
Murakami, T., Tanaka, Y., Kulkarni, S.~R., Ogasaka, Y., Sonobe, T., Ogawara,
  Y., Aoki, T., \& Yoshida, T. 1994, { Nature}, {\rm 368}, 127.

\bibitem[Nicastro, Johnston, \& Koribalski 1996]{njk96}
Nicastro, L., Johnston, S., \& Koribalski, B. 1996, { A\&A}, {\rm 306}, 49.

\bibitem[Paczy\'{n}ski 1992]{pac92}
Paczy\'{n}ski, B. 1992, { Acta Astron.}, {\rm 42}, 145.

\bibitem[Retallack 1983]{ret83}
Retallack, D.~S. 1983, { MNRAS}, {\rm 204}, 669.

\bibitem[Rho \& Petre 1997]{rp97}
Rho, J. \& Petre, R. 1997, { ApJ}, {\rm 484}, 828.

\bibitem[Rothschild, Kulkarni, \& Lingenfelter 1994]{rkl94}
Rothschild, R.~E., Kulkarni, S.~R., \& Lingenfelter, R.~E. 1994, { Nature},
  {\rm 368}, 432.

\bibitem[Sanbonmatsu \& Helfand 1992]{sh92b}
Sanbonmatsu, K.~Y. \& Helfand, D.~J. 1992, { AJ}, {\rm 104}, 2189.

\bibitem[Sarma \etal  1997]{sggf97}
Sarma, A.~P., Goss, W.~M., Green, A.~J., \& Frail, D.~A. 1997, { ApJ}, {\rm
  483}, 335.

\bibitem[Sault, Staveley-Smith, \& Brouw 1996]{ssb96}
Sault, R.~J., Staveley-Smith, L., \& Brouw, W.~N. 1996, { A\&AS}, {\rm 120},
  375.

\bibitem[Sault \& Wieringa 1994]{sw94}
Sault, R.~J. \& Wieringa, M.~H. 1994, { A\&AS}, {\rm 108}, 585.

\bibitem[Schwentker 1994]{sch94}
Schwentker, O. 1994, { A\&A}, {\rm 286}, L47.

\bibitem[Shaver \& Goss 1970]{sg70c}
Shaver, P.~A. \& Goss, W.~M. 1970, { Aust. J. Phys. Astr. Supp.}, {\rm 14},
  133.

\bibitem[{Shitov}, {Pugachev}, \& {Kutuzov} 2000]{spk00}
{Shitov}, Y.~P., {Pugachev}, V.~D., \& {Kutuzov}, S.~M. 2000, in { Pulsar
  Astronomy --- 2000 and Beyond, {IAU} Colloquium 177}, ed.\ M. Kramer, N. Wex,
  \& R. Wielebinski, (San Francisco: Astronomical Society of the Pacific), 685.

\bibitem[Shull, Fesen, \& Saken 1989]{sfs89}
Shull, J.~M., Fesen, R.~A., \& Saken, J.~M. 1989, { ApJ}, {\rm 346}, 860.

\bibitem[Shull~Jr. 1983]{shu83}
Shull~Jr., P. 1983, { ApJ}, {\rm 275}, 611.

\bibitem[Slane \etal  2000]{scs+00}
Slane, P., Chen, Y., Schulz, N.~S., Seward, F.~D., Hughes, J.~P., \& Gaensler,
  B.~M. 2000, { ApJ}, {\rm 533}, L29.

\bibitem[Smith, Bradt, \& Levine 1999]{sbl99}
Smith, D.~A., Bradt, H.~V., \& Levine, A.~M. 1999, { ApJ}, {\rm 519}, L147.

\bibitem[Smith \etal  1994]{scm+94}
Smith, R.~C., Chu, Y.-H., MacLow, M.-M., Oey, M.~S., \& Klein, U. 1994, { AJ},
  {\rm 108}, 1266.

\bibitem[Song \etal  2000]{smm+00}
Song, L., Mihara, T., Matsuoka, M., Negoro, H., \& Corbet, R. 2000, { PASJ},
  {\rm 52}, 181.

\bibitem[Stappers, Gaensler, \& Johnston 1999]{sgj99}
Stappers, B.~W., Gaensler, B.~M., \& Johnston, S. 1999, { MNRAS}, {\rm 308},
  609.

\bibitem[{Sugizaki} \etal  1997]{snt+97}
{Sugizaki}, M., {Nagase}, F., {Torii}, K.~I., {Kinugasa}, K., {Asanuma}, T.,
  {Matsuzaki}, K., {Koyama}, K., \& {Yamauchi}, S. 1997, { PASJ}, {\rm 49},
  L25.

\bibitem[Tateyama, Strauss, \& Kaufmann 1991]{tsk91}
Tateyama, C.~E., Strauss, F.~M., \& Kaufmann, P. 1991, { MNRAS}, {\rm 249},
  716.

\bibitem[Tauris \& Manchester 1998]{tm98}
Tauris, T.~M. \& Manchester, R.~N. 1998, { MNRAS}, {\rm 298}, 625.

\bibitem[Terrell \etal  1980]{tekl80}
Terrell, J., Evans, W.~D., Klebesadel, R.~W., \& Laros, J.~G. 1980, { Nature},
  {\rm 285}, 383.

\bibitem[Thompson \& Blaes 1998]{tb98}
Thompson, C. \& Blaes, O. 1998, { Phys. Rev. D}, {\rm 57}, 3219.

\bibitem[Thompson \& Duncan 1993]{td93a}
Thompson, C. \& Duncan, R.~C. 1993, { ApJ}, {\rm 408}, 194.

\bibitem[{Thompson} \& {Duncan} 1995]{td95}
{Thompson}, C. \& {Duncan}, R.~C. 1995, { MNRAS}, {\rm 275}, 255.

\bibitem[Thompson \& Duncan 1996]{td96b}
Thompson, C. \& Duncan, R.~C. 1996, { ApJ}, {\rm 473}, 322.

\bibitem[{Thompson} \etal  2000]{tdw+00}
{Thompson}, C., {Duncan}, R.~C., {Woods}, P.~M., {Kouveliotou}, C., {Finger},
  M.~H., \& {van Paradijs}, J. 2000, { ApJ}, {\rm 543}, 340.

\bibitem[Torii \etal  1998]{tkk+98}
Torii, K., Kinugasa, K., Katayama, K., Tsunemi, H., \& Yamauchi, S. 1998, {
  ApJ}, {\rm 503}, 843.

\bibitem[Tovmassian \etal  1998]{tgct98}
Tovmassian, H.~M., Gonzalez, A., Corral, L.~J., \& Tovmassian, G, H. 1998, {
  Astrophys. Space Sci.}, {\rm 257}, 63.

\bibitem[van Kerkwijk \etal  1995]{vkmn95}
van Kerkwijk, M.~H., Kulkarni, S.~R., Matthews, K., \& Neugebauer, G. 1995, {
  ApJ}, {\rm 444}, L33.

\bibitem[van Paradijs, Taam, \& van~den Heuvel 1995]{vtv95}
van Paradijs, J., Taam, R.~E., \& van~den Heuvel, E. P.~J. 1995, { A\&A}, {\rm
  299}, L41.

\bibitem[Vancura \etal  1992]{vblr92}
Vancura, O., Blair, W.~P., Long, K.~S., \& Raymond, J.~C. 1992, { ApJ}, {\rm
  394}, 158.

\bibitem[Vasisht, Frail, \& Kulkarni 1995]{vfk95}
Vasisht, G., Frail, D.~A., \& Kulkarni, S.~R. 1995, { ApJ}, {\rm 440}, L65.

\bibitem[Vasisht \& Gotthelf 1997]{vg97}
Vasisht, G. \& Gotthelf, E.~V. 1997, { ApJ}, {\rm 486}, L129.

\bibitem[Vasisht \etal  2000]{vgtg00}
Vasisht, G., Gotthelf, E.~V., Torii, K., \& Gaensler, B.~M. 2000, { ApJ}, {\rm
  542}, L49.

\bibitem[{Vrba} \etal  2000]{vhl+00}
{Vrba}, F.~J., {Henden}, A.~A., {Luginbuhl}, C.~B., {Guetter}, H.~H.,
  {Hartmann}, D.~H., \& {Klose}, S. 2000, { ApJ}, {\rm 533}, L17.

\bibitem[Walsh \etal  1998]{wbhr98}
Walsh, A.~J., Burton, M.~G., Hyland, A.~R., \& Robinson, G. 1998, { MNRAS},
  {\rm 301}, 640.

\bibitem[Weinberg \& Nikolaev 2001]{wn01}
Weinberg, M.~D. \& Nikolaev, S. 2001, { ApJ}, {\rm 548}, 712.

\bibitem[{White} \etal  1996]{wae+96}
{White}, N.~E., {Angelini}, L., {Ebisawa}, K., {Tanaka}, Y., \& {Ghosh}, P.
  1996, { ApJ}, {\rm 463}, L83.

\bibitem[Whiteoak 1994]{whi94}
Whiteoak, J. B.~Z. 1994, { ApJ}, {\rm 429}, 225.

\bibitem[Whiteoak \& Green 1996]{wg96}
Whiteoak, J. B.~Z. \& Green, A.~J. 1996, { A\&AS}, {\rm 118}, 329.
\newblock (http://www.physics.usyd.edu.au/astrop/wg96cat/).

\bibitem[Wiliams \etal  1999]{wcd+99}
Wiliams, R.~M., Chu, Y.-H., Dickel, J.~R., Petre, R., C., S.~R., \& Tavarez, M.
  1999, { ApJS}, {\rm 123}, 467.

\bibitem[Williams \etal  1997]{wcd+97}
Williams, R.~M., Chu, Y.-H., Dickel, J.~R., Beyer, R., Petre, R., Smith, R.~C.,
  \& Milne, D.~K. 1997, { ApJ}, {\rm 480}, 618.

\bibitem[{Woods} \etal  1999a]{wkv+99b}
{Woods}, P.~M. \etal  1999a, { ApJ}, {\rm 524}, L55.

\bibitem[{Woods} \etal  1999b]{wkv+99}
{Woods}, P.~M. \etal  1999b, { ApJ}, {\rm 519}, L139.

\bibitem[{Young}, {Manchester}, \& {Johnston} 1999]{ymj99}
{Young}, M.~D., {Manchester}, R.~N., \& {Johnston}, S. 1999, { Nature}, {\rm
  400}, 848.

\bibitem[Zhang, Xu, \& Qiao 2000]{zxq00}
Zhang, B., Xu, R.~X., \& Qiao, G.~J. 2000, { ApJ}, {\rm 545}, L127.

\bibitem[Zoonematkermani \etal  1990]{zhb+90}
Zoonematkermani, S., Helfand, D.~J., Becker, R.~H., White, R.~L., \& Perley,
  R.~A. 1990, { ApJS}, {\rm 74}, 181.

\end{thebibliography}

\clearpage

\begin{table}[hbt]
\vspace{3cm}
\caption{Summary of Observations}
\begin{tabular}{lccccc} \hline \hline
Source   & Telescope & Date of  & Array & Observing & Time on \\ 
         &           & Observations   & Configuration & Frequency (GHz) 
	 & Source (h) \\ \hline
\rxj\   & VLA  & 2000 Mar 09 & CnB & 1.4 & 1.3 \\
	&      & 2000 Jul 08 & DnC & 1.4 &  1.0 \\  \hline
\4u\    & VLA  & 2000 Mar 23 & C   & 1.4 & 1.0 \\
	&      & 2000 May 09 & C   & 1.4 & 1.5 \\ \hline \hline
\end{tabular}
\label{tab_obs}
\end{table}

\begin{table}[hbt]
\vspace{3cm}
\caption{Claimed associations of SNRs with AXPs and SGRs.}
\label{tab_snrs}
\begin{tabular}{llccccccl} \tableline \tableline
AXP or SGR   & SNR  &  $t_{\rm SNR}$ & $d_{\rm SNR}$ & $\Delta\theta$ & 
$\theta_{\rm SNR}$  & $\beta$ & $V_T$ & Reference \\
         &      &  (kyr)          & (kpc) &  (arcmin) & (arcmin)  & &
	 (\kms)  \\ \tableline
1E~1841--045 & Kes~73 & 2 & 7 & $<$0.5 & 2 & $<$0.25 & $<$500  & 1, 2 \\
AX~J1845--0258 & G29.6+0.1 & $<$8 & $<$20 &$<$0.65 & 2.5 & 
$<$0.25 & $<$500 & 3, 4 \\
1E~2259+586 & CTB~109 & $\sim$10 & 5 & $<$3 & 14 & $<$0.2 & $<$400 & 5, 6  \\ \tableline
SGR~0526--66 & N49 & $5-16$ & 50 & 0.5 & 0.5 & 1 & 400--1400 &   7, 8 \\
\1627\ & G337.0--0.1 &  $1-5$ & 11  & 1.3 & 0.75 & 1.7 & 800--4000 & 9 \\
SGR~1900+14 & G42.8+0.6 & --- & --- & 17 & 12 & 1.4 & --- & --- \\
\tableline \tableline
\end{tabular}

References for
ages and distances:
(1) Sanbonmatsu \& Helfand (1992\protect\nocite{sh92b});
(2) Vasisht \& Gotthelf (1997\protect\nocite{vg97}); (3) Gaensler \etal\
(1999\protect\nocite{ggv99}); 
(4) Vasisht \etal\ (2000\protect\nocite{vgtg00});
(5) Green (1989\protect\nocite{gre89});
(6) Rho \& Petre (1997\protect\nocite{rp97}); 
(7) Shull (1983\nocite{shu83});
(8) Vancura \etal\ (1992\protect\nocite{vblr92}); 
(9) Corbel \etal\ (1999\protect\nocite{ccdd99}).
\end{table}

\clearpage

\clearpage

\clearpage

\end{document}